\documentclass[reprint,english,aps,prl,twocolumn, superscriptaddress]{revtex4}
\usepackage[T1]{fontenc}
\usepackage{subfigure}
\usepackage[latin1]{inputenc}
\usepackage{graphicx}
\usepackage{epsfig}
\usepackage{prettyref}
\usepackage{babel}
\usepackage{caption}
\makeatother

\begin{document}

\title{Experimental evidence for stochastic switching of supercooled phases}

\author{Devendra Kumar}
\email{deveniit@gmail.com}
\affiliation{Department of Physics, Indian Institute of Technology Kanpur 208016, India}
\affiliation{UGC-DAE Consortium for Scientific Research, University Campus, Khandwa Road, Indore-452001, India}

\author{K. P. Rajeev}
\email{kpraj@iitk.ac.in}\affiliation{Department of Physics, Indian
Institute of Technology Kanpur 208016, India}

\author{J. A. Alonso}
\affiliation{Instituto de Ciencia de Materiales de Madrid, CSIC,
Cantoblanco, E-28049 Madrid, Spain}

\author{M. J. Martínez-Lope}
\affiliation{Instituto de Ciencia de Materiales de Madrid, CSIC,
Cantoblanco, E-28049 Madrid, Spain}

\begin{abstract}
First-order phase transition in a highly correlated electron system can manifest as a dynamic phenomenon.
The presence of multiple domains of the coexisting phases average out the dynamical effects making it nearly impossible to predict the exact nature of phase transition dynamics. Here we report the metal-insulator transition in samples of sub-micrometer size NdNiO$_3$ where the effect of averaging is minimized by restricting the number of domains under study. We observe the presence of supercooled metallic phases with supercooling of 40~K or more. The transformation from supercooled metallic to insulating state is a stochastic process that happens at different temperature and time in different experimental runs. The experimental results are understood without incorporating material specific properties suggesting their universal nature. The size of the sample needed to observe individual switching of supercooled domains, the degree of supercooling, and the time-temperature window of switching is expected to depend on the parameters such as quenched disorder, strain, magnetic field etc.
\end{abstract}

\maketitle

The inherent dynamics of a first order phase transition often manifests itself in the form of time dependence in physical properties in a number of transition metal oxides, for example, in VO$_2$\cite{Morin, Claassen}, CMR manganites\cite{Wu, Levy, Ward, Ghivelder}, and nickelates\cite{Granados, Devendra}. Recently a number of attempts have been made to get an insight into the mechanism of phase transition by limiting the number of domains under study, which reduces the averaging that occurs due to phase transition in multiple domains during the time scale of measurement, and therefore, enables the bulk measurements like resistivity to probe the effect of the transition in individual domains. The resistivity measurements on (La$_{5/8-0.3}$Pr$_{0.3}$)Ca$_{3/8}$MnO$_3$ (LPCMO) nanowires show giant discrete steps in the metal to insulator transition regime, the size and positions of these jumps vary in different thermal runs\cite{Zhai}. Similar jumps in resistivity have also been observed in VO$_2$ thin films\cite{Kawatani}, nanostrucutres\cite{Sharoni} and nanobeams\cite{Wei1}. The existence of multiple jumps in temperature scan of resistivity are attributed to percolative transport in the limited number of domains that have different transition temperature due to quenched disorder\cite{Zhai, Sharoni}, but the random change in the position and size of the jumps can not be understood only on the basis of this proposition. The typical variation in the temperatures of resistance jump in different thermal scans is around 3-8~K in LPCMO nanowires\cite{Zhai}, less than 1~K in VO$_2$ nanostructures\cite{Sharoni}, and around 2-5~K in suspended nanobeams of VO$_2$\cite{Wei1}. The size of resistivity jumps are observed to depend on strain\cite{Kawatani, Wei1, Zhang, Wu2, Cao}, magnetic field\cite{Zhai}, and device length\cite{Zhai, Sharoni}. A deeper understanding of the dynamics of phase transformation process in the correlated electron system is required to answer whether occurrence of multiple jumps at different temperatures in different thermal runs is a generic property of the first order phase transition or is a material specific property. If it is a generic property then the question of interest would be what microscopic mechanism is responsible for such events, how these events are connected with dynamical properties observed at macroscopic length scale, and how they depend on the parameters that control the phase transition in general.

The rare earth nickelate NdNiO$_3$ undergoes a first order metal-insulator transition at 200~K\cite{Granados1, Medarde}. This metal-insulator transition has a large thermal hysteresis, and in the hysteresis regime, NdNiO$_3$ exhibits time dependence in physical properties\cite{Devendra} which makes it one of the suitable candidate to probe the universality of the occurrence of random resistance jumps and their possible implication on the dynamical effects at macroscopic scale. We restrict the number of domains under study by depositing micron size silver electrodes at sub-micron separation. These electrodes are deposited on an NdNiO$_3$ pellet that was earlier used in a time dependent study reported in Ref.~\onlinecite{Devendra} to ensure that the sample in between the silver electrodes have exactly the same microstructure as that used in the earlier studies. This is important because the dynamical properties across a phase transition depend significantly on sample microstructure such as density of defects, crystallite size, strain etc\cite{Devendra2}.

\begin{figure}[]
\begin{centering}
\includegraphics[width=0.5\columnwidth]{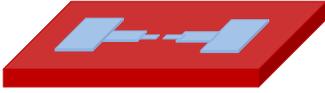}
\par\end{centering}
\caption{Schematic of contact electrodes. The red region represents NdNiO$_3$ pellet while the blue represents silver contact electrodes.}
\label{fig: nano}
\end{figure}

\begin{figure}[]
\begin{centering}
\includegraphics[width=0.6\columnwidth]{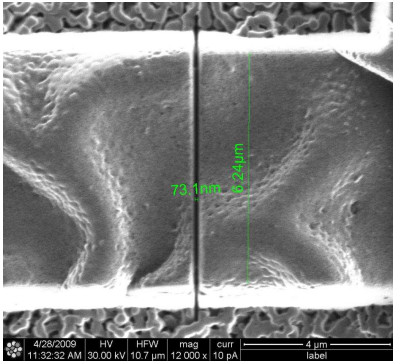}
\par\end{centering}
\caption{Scanning electron micrograph of 6~$\mu$m wide electrodes fabricated at around 70~nm separation on the polycrystalline NdNiO$_3$ pellet.}
\label{fig: SEM}
\end{figure}

\begin{figure}[t]
\begin{centering}
\includegraphics[width=0.7\columnwidth]{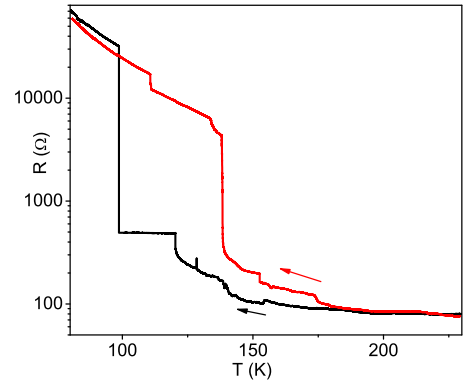}
\par\end{centering}
\caption{Resistance versus temperature curves for the 6~$\mu$m $\times$ 70~nm sample in two consecutive cooling runs. The black curve represents the first cooling run while the red curve represents the immediately following cooling run.} 
\label{fig: RvsT80nm}
\end{figure}

Two silver contact pads of 1.5 mm$\times$ 1.5 mm connected via a 100~$\mu$m wide silver strip are sputter deposited on the polished surface of NdNiO$_3$ using shadow masking. Schematic diagram is shown in Fig.~\ref{fig: nano}. The 100~$\mu$m silver strip is ion milled using a focussed ion beam (Nova 600 Nano Lab FIB) to fabricate two 6~$\mu$m wide silver electrodes separated by distance of 70~nm. A similar device with silver electrode separation of 120~nm is also fabricated. The SEM image of 6~$\mu$m wide contact electrodes at 70~nm separation is shown in Fig.~\ref{fig: SEM}. The resistance of these samples are measured by quasi-four probe method and the rate of temperature ramp is fixed at 0.2~K/min. The results of resistivity measurements performed on the sample with 6~$\mu$m contact pads at 70~nm separation in two consecutive cooling runs are displayed in Fig.~\ref{fig: RvsT80nm}. The resistance versus temperature curve for the two cooling runs coincides above 200~K but follow different paths on cooling below 200~K. On lowering the temperature below 200~K the resistance exhibits an overall growth indicating the occurrence of metal to insulator transition in the domains confined between the silver electrodes, but this growth is not smooth as in the case of bulk samples\cite{Devendra}.  The initial growth in resistance is accompanied by tiny jumps and dips, and below 155~K, the resistance jumps  become sharp with a maximum value of around 30~k$\Omega$.

The typical crystallite size of NdNiO$_3$  pellet is larger than 100~nm, and therefore many of the crystallites in between the contact pads form a direct bridge between the pads. In this configuration, the temperature of 
highest order resistance jump exhibits the metal to insulator transition in the last metallic crystallite directly bridging the  pads\cite{endnote}. The sharp jumps in resistance occurs at different temperatures in consecutive cooling runs, for example, the highest order resistance jump occurs around 98~K and 138~K in the first and second cooling run respectively. See Fig.~\ref{fig: RvsT80nm}. This difference in the temperature of highest order resistance jump shows that the last metallic crystallite bridging the contact pads undergoes a metal to insulator transformation at the temperature difference of 40~K in the two consecutive runs. After two thermal cycles the  silver contact pads degrade and the data can not be recorded for further cycles.
Below 100~K, the resistance follows the behavior of a band-gap insulator with data of both runs almost falling on the same curve.

\begin{figure}[]
\begin{centering}
\includegraphics[width=0.8\columnwidth]{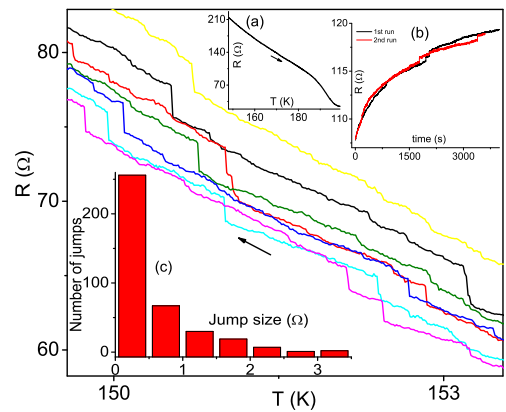}
\par\end{centering}
\caption{Resistance versus temperature curves  for 6~$\mu$m $\times$ 120~nm sample in seven consecutive cooling runs. Each cooling run is shown in a different colour. The inset (a) shows the temperature variation of resistance in one of the heating runs. The inset (b) shows the time evolution of resistance at 145~K in two consecutive cooling runs. The inset (c) displays the size distribution of resistance jumps observed in resistance versus temperature data from all the seven consecutive cooling runs}
\label{fig: RvsT120nm}
\end{figure}

Similar set of resistivity measurements are also performed with 6~$\mu$m silver contact pads milled at 120~nm separation. These contact pads are relatively more stable for thermal cycling. The results of resistance versus temperature measurement up to seven consecutive cooling runs
are presented in Fig.~\ref{fig: RvsT120nm}. The resistance shows multiple jumps of different size on approaching the percolation threshold in the cooling runs, but as shown in inset (a) of Fig.~\ref{fig: RvsT120nm}, no such resistance jumps are observed in the heating run. The temperature of these resistance jumps vary randomly on each cooling run which indicates the stochastic nature of the underlying physical mechanism. This is also confirmed by the results of time dependent resistance measurements (see inset (b) of Fig.~\ref{fig: RvsT120nm}) where small jumps in resistance are observed at different times in two consecutive cooling runs. The histogram in inset (c) of Fig.~\ref{fig: RvsT120nm} displays the size distribution of resistance jumps collected from resistance versus temperature measurements over seven cooling runs. The number of jumps decreases rapidly as the jump size increases.
The size of resistance jumps for 120~nm contact pad separation are much smaller than that seen in case of 70~nm, and in contrast to 70~nm sample, these jumps are visible only around the percolation regime. This suggests that the current transport mechanism alters drastically on going from 70~nm  to 120~nm contact pad separation. For 70 nm separation most of the crystallites directly bridge the contact electrodes making the current flow similar to a parallel resistor network, but for 120~nm, the number of non bridging crystallites dominate which shifts the current flow mechanism from the parallel resistor network to a complex combination of series-parallel resistors which can be better analyzed by effective medium theory\cite{McLachlan}.

\begin{figure}[!t]
\begin{centering}
\includegraphics[width=0.8\columnwidth]{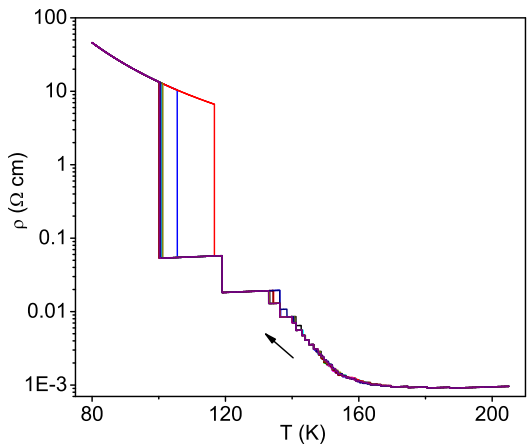}
\par\end{centering}
\caption{Simulation results of 50 coherence volumes sample in different cooling runs. Each cooling run is shown by a different color. The volume to resistivity conversion was done using parallel resistor networks.}
\label{fig: 50 SR}
\end{figure}

The resistivity measurements on the same NdNiO$_3$ pellet, but with large number of crystallites have shown that the metal insulator transition is broadened due to the presence of quenched disorder that creates a local variation in the transition temperature. The dynamical measurements in the hysteresis regime show an increase in insulating volume fraction with time at the cost of metallic phases. These results are understood on the basis of the conjecture that the high temperature metallic phase can exist in a supercooled metallic state below its metal insulator transition temperature and the increase in insulating volume fraction with time is an outcome of stochastic switching of metastable supercooled regions into stable insulating state\cite{Devendra}. A supercooled metallic region undergoing switching as a single entity can be a crystallite or a part of it and will be termed as a 'coherence volume' in the subsequent discussion\cite{Endnote}. The switching of a supercooled metallic coherence volume to insulating state is a two step process: The first step involves the nucleation of a stable critical nuclei of insulating phase inside the supercooled region while the second step involves the growth of critical nuclei\cite{Lifshitz}, and if the kinetics of the constituent atoms of the material is not arrested then the growth of the critical nuclei is a relatively fast process with typical time scale of the order of $\mu$s\cite{Sharma}. Therefore the timescale of the switching of a supercooled metallic phase to stable insulating state is mainly determined by the time required for the formation of critical nuclei.

\begin{figure}[!t]
\begin{centering}
\includegraphics[width=0.8\columnwidth]{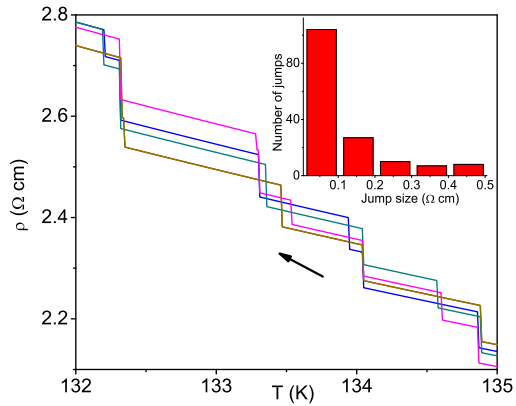}
\par\end{centering}
\caption{Simulation results of 500 coherence volumes sample in different cooling runs. Each cooling run is shown by a different color. The volume to resistivity conversion was done using the GEM equation. The inset shows the size distribution of resistance jumps taken from simulated resistance versus temperature curves of the main panel.}
\label{fig: 500 SR}
\end{figure}

For 70~nm electrode separation, our results show that in two consecutive cooling runs, the last metallic coherence volume bridging the contact electrode switches from metallic to insulating state at 98~K and 138~K respectively. The coherence volume switching at 98~K and 138~K may be same or different, but in both the cases, the metal insulator transition temperature ($T_{MI}$) of the coherence volume should be equal to or above 138~K. This is because if the switching in first run occurs at the $T_{MI}$ then it should occur at $T_{MI}$ or below in both of the cooling runs. The existence of the coherence volume in the metallic state up to 98~K, which is below its $T_{MI}$, gives a clear evidence of the existence of metallic phases in the supercooled state with 
supercooling of 40~K or more. The results of 120~nm electrode separation suggest that underlying physical mechanism of resistance jumps is stochastic in nature. A jump in resistance is expected when a metallic coherence volume switches to the insulating state.  If this switching occurs at $T_{MI}$ then the jump should occur at the same temperature in different cooling runs, but the random change in temperature of jumps in different cooling runs show that the metallic phases exist in the supercooled state with a varying degree of supercooling. The arbitrary variation in the degree of supercooling along with the randomness in time taken for switching from supercooled to ground state at a given temperature in different cooling runs suggest that the transformation from supercooled to stable state is a stochastic process.

The universality of the random resistance jumps in the percolation regime of sub-micrometer size first order transition systems can be established if the existence of resistance jumps can be attributed to the stochastic switching of supercooled coherence volumes invoking only minimal material specific properties. The idea of the universality of random resistance jumps is put to test by calculating the resistivity of a representative set of limited number of coherence volumes which mimic a sub-micrometer size first order transition system.
We generate such a set of 50 and 500 'coherence volumes' whose temperature of the limit of metastability ($T^*$) and volume are distributed similar to that of bulk NdNiO$_3$. The resistivity of these set of coherence volumes is calculated by performing simulation on the basis of the model used for calculating time dependent effects in Ref.~\onlinecite{Devendra}. The model states that the supercooled metallic coherence volumes can switch to the insulating state stochastically anywhere between $T_{MI}$ and $T^*$. The probability of transformation ($p$) is proportional to exp$(-U/k_BT)$ where U is the energy required for the formation of critical nuclei of the insulating phase. $U$ is maximum at $T_{MI}$, which decreases on lowering the temperature, and vanishes at $T^*$, and therefore the probability of transformation increases as $T$ approaches $T^*$. In different experimental runs, a  supercooled metallic coherence volume may transform from the metallic to the insulating state at different temperatures. The resistivity of the 50 and 500 coherence volume samples are calculated using the parallel resistor network and effective medium theory respectively. See supplementary material for the details of the calculation. The resistivity results of the simulation are shown in Fig.~\ref{fig: 50 SR} and~\ref{fig: 500 SR} and the pattern of temperature dependence of resistivity for 50 and 500 coherence volumes in different cooling runs replicate the experimental results of 70~nm and 120~nm electrode separation respectively. A comparison of the size distribution of resistance jumps for 500 coherence volume (inset of Fig.~\ref{fig: 500 SR}) with that of 120~nm electrode separation sample (inset (c) of Fig.~\ref{fig: RvsT120nm}) show that they exhibit very similar trends. Since no material specific property except the distribution of $T^*$, coherence volume size, and energy of nucleation is used in these calculations, it shows that the existence of the random resistance jumps in resistivity or in general, the existence of random jumps in any other physical property is the characteristic of any first order transition system measured at the crystallite size scale.

A comparison of the results of Fig.~\ref{fig: 50 SR} and~\ref{fig: 500 SR} show that an enhancement in the number of coherence volumes, where each coherence volume has its own T$_{MI}$,  reduces the size of resistivity jumps. The presence of quenched disorder spreads the local transition temperature which in turn reduces the size of coherence volume, and therefore, it effectively enhances the number of coherence volumes in a given system. This will reduce the size of resistance jumps. The maximum allowed variation in the temperature of the resistance jump due to switching of a particular coherence volume in different cooling runs is $T_{MI}-T^*$, where $T_{MI}$ is the transition temperature and $T^*$ is the limit of metastability of that volume. The presence of quenched disorder also reduces the energy of nucleation of the stable phase inside the metastable phase and this in turn reduces the possible degree of supercooling ($T_{MI}-T^*$) i.e. the temperature range in which a supercooled metastable phase can exist\cite{Devendra2}. Therefore the nanoscale samples with enhanced quenched disorder will exhibit a small variation in temperature of resistance jumps in different cooling runs and the size of these resistance jumps will also be small.

In resistivity measurements of LPCMO nano-wires\cite{Zhai}, the size of resistivity jumps and the thermal hysteresis in resistivity decreases on increasing the magnetic field and finally vanishes at around 6~T. The decay in magnitude of the resistivity jump suggests that possibly the size of coherence volume decreases on cooling at higher fields in LPCMO nano-wires while the decay in thermal hysteresis indicates that the degree of supercooling of the coherence volume also decreases on cooling at higher fields. In the process of phase transformation, the size of coherence volume within a crystallite is determined by (a) spatial variation in free energy density due to variation in local density of quenched disorder, (b) energy released due to difference in free energy of metallic and insulating states and (c) energy required for the formation of interface of correlated volume\cite{Imry}. If the applied magnetic field (or pressure etc.) enhances the difference in free energy density of supercooled metastable and stable phases, which is the case for LPCMO nano-wires where applied magnetic field reduces the free energy density of stable ferromagnetic metallic phases much more than that of supercooled charge ordered antiferromagnetic insulating phases, the size of coherence volume reduces as the energy is available to form interfaces as favoured by the spatial variation in the free energy density. An enhancement in the size of coherence volume and resistance jumps is expected if the applied field reduces the difference in free energy density of supercooled and stable phases.

In summary, the observation of random resistance jumps in the nano-structures of the first order transition system establishes that the switching from supercooled to stable state is a stochastic process. The dynamical properties observed in first order systems originate from such stochastic switching. The degree of supercooling, the probability of switching, the coherence volume of supercooled phases, and their number in a given sample depends on the parameters such as density of defects, magnetic field, pressure etc and therefore the size and temperature variation of the random resistance jumps can be tuned by these variables which opens up the possibility of engineering these nano-structures for functional purposes.

\section{Supplementary Material}
\subsection{Simulation details}
\begin{figure}[]
\begin{centering}
\includegraphics[width=0.6\columnwidth]{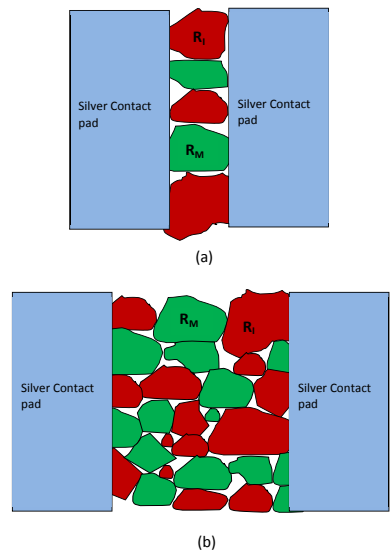}
\par\end{centering}
\caption{Schematic of coherence volumes and contact pads for different sample sizes. (a) When the sample size is small most of the coherence volume make a bridge between the contact electrodes. The equivalent resistance can be calculated from parallel resistor formula. When the sample is relatively large the situation is as depicted in (b). In this case the GEM equation is a more appropriate choice.}
\label{fig: mixed phase}
\end{figure}

Two sets of 50 and 500 number of coherence volume are generated with volume ratio uniformly distributed between 1 and 6. In each set, the coherence volumes are assigned a $T^*$ in such a way that the temperature dependence of the volume distribution of supercooled metallic coherence volumes follow the pattern of the bulk sample given in Ref.~\onlinecite{Devendra}.
The temperature variation of insulating volume fraction in the cooling run (at a given cooling rate) is calculated by lowering the temperature in step of $\Delta T$ and switching the supercooled metallic coherence volumes having $T^*$ in between $T$ and $T-\Delta T$ to the insulating state. This is followed by a time step $\Delta t$ during which all the remaining supercooled metallic coherence volumes are given a chance to switch to insulating state isothermally. The details of the method and the parameters used in simulation are described in Ref.~\onlinecite{Devendra}. For small samples, the connecting electrodes are very close to each other and the coherence volume will form bridges between the electrodes as shown in Fig.~\ref{fig: mixed phase}~(a).  The resistivity of such a nano-scale sample can be calculated using a  parallel resistor formula which can be written as,
\begin{equation}
\frac{1}{\rho}= \frac{f_M}{\rho_M} +  \frac{f_I}{\rho_I},
\label{eq:resistivity nano}
\end{equation}
where $f_M$ is the metallic volume fraction, $f_I$ is the insulating volume fraction, $\rho_M$ is the resistivity of the metallic regions, and $\rho_I$ is the resistivity of the insulating regions. When the size of the sample is relatively large, there may be multiple coherence volumes between the contact electrodes of the nanostructure (see Fig.~\ref{fig: mixed phase} (b)). These coherence volumes can be either in metallic or insulating state and therefore the system between the electrodes can be modelled as a binary macroscopic mixture of two different electrical resistivities. In this case, McLachlan's general effective medium (GEM) equation becomes a  more appropriate choice for volume fraction to resistivity conversion\cite{McLachlan}. This is illustrated in the following paragraph.

\begin{figure}[]
\begin{centering}
\includegraphics[width=0.7\columnwidth]{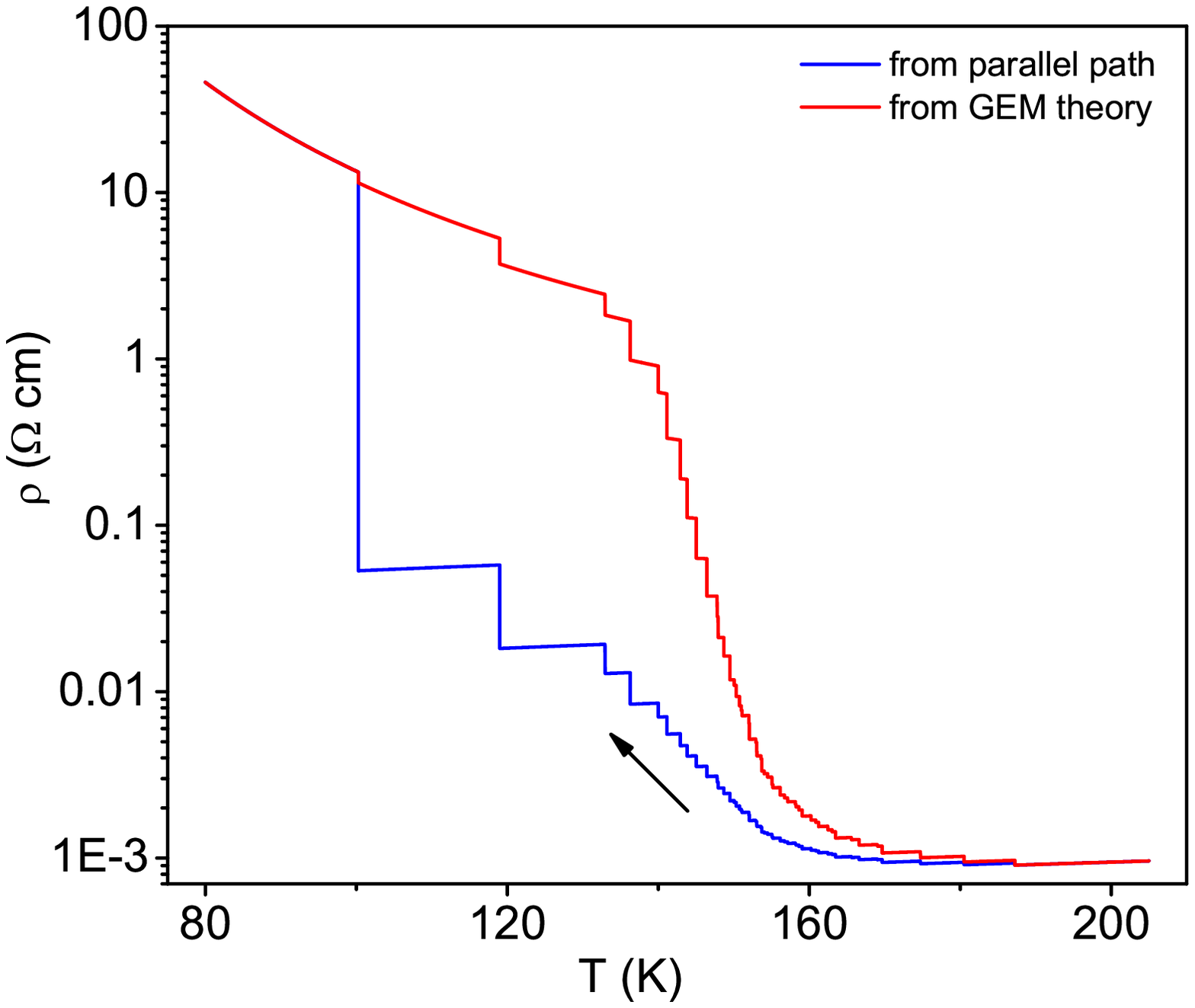}
\par\end{centering}
\caption{Simulation results for the 50 coherence volumes sample in a cooling run. The blue line shows the resistivity obtained from the parallel resistor network while the red line shows the resistivity obtained using the GEM equation.}
\label{fig: GEM vs parallel 50SR}
\end{figure}

\begin{figure}[!t]
\begin{centering}
\includegraphics[width=0.7\columnwidth]{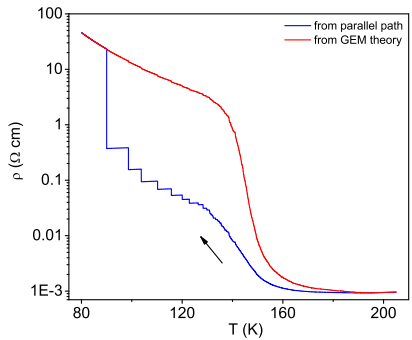}
\par\end{centering}
\caption{Simulation results for the 500 coherence volumes sample in a cooling run. The blue line shows the resistivity obtained from the parallel resistor network while the red line shows the resistivity obtained from the GEM equation.}
\label{fig: GEM vs parallel 500SR}
\end{figure}

Fig.~\ref{fig: GEM vs parallel 50SR} shows the simulation results of the 50 coherence volume sample. We can see that when parallel resistor network is used for volume to resistivity conversion, the last few jumps in resistivity are much bigger than the others. These jumps are expected because the last few metallic connections between the contact electrodes break at these points. These big jumps are preceded by flat plateaus in resistivity because in the plateau region the change in volume is zero and the resistivity is determined by the last few metallic coherence volumes bridging the contact electrodes. When GEM equation is used for volume to resistivity conversion the resistivity jumps are of nearly same size. In Fig.~\ref{fig: GEM vs parallel 500SR} we show the simulation results for 500~coherence volume sample. Here resistivity obtained by parallel resistor network has large jumps similar to that of 50 coherence volume sample but the resistivity obtained through GEM equation is nearly smooth. Since the size of the resistivity jumps in experiments are large when sample is small and vanishes on approaching the bulk, the parallel path network is more appropriate for volume to resistivity conversion for the 50 coherence volume set while the GEM equation is more suited for that of 500 coherence volume set. The similar trend in size distribution of resistance jumps in 120~nm electrode separation sample (inset (c) of the Fig.~4 of the manuscript) and 500 coherence volume sample (inset of Fig.~6 of the manuscript) indicates that the GEM theory gives a fairly accurate result for volume to resistivity conversion of 500 coherence volume.

\section{Acknowledgements}
DK and KPR are thankful to R. C. Budhani for useful discussions. DK thanks the University Grants Commission of India for financial
support. JAA and MJM-L acknowledge the Spanish Ministry of
Education for funding the Project MAT2010-16404.


\begin{thebibliography}{20}

\bibitem{Morin} F. J. Morin, Phys. Rev. Lett. \textbf{3}, 34 (1959).

\bibitem{Claassen} J. H. Claassen, J. W. Lu, K. G. West, and S. A. Wolf, Appl. Phys. Lett. \textbf{96}, 132102 (2010).

\bibitem{Wu} W. Wu, C. Israel, N. Hur, S. Park, S. -W. Cheong, and A. Lozanne, Nature Materials \textbf{5}, 881 (2006).

\bibitem{Levy} P. Levy,  F. Parisi, L. Granja, E. Indelicato, and  G. Polla, Phys. Rev. Lett. \textbf{89}, 137001 (2002).

\bibitem{Ward} T. Z. Ward, Z. Gai, H. W. Guo, L. F. Yin, and J. Shen, Phys. Rev. B \textbf{83}, 125125 (2011).

\bibitem{Ghivelder} L. Ghivelder and F. Parisi, Phys. Rev. B \textbf{71}, 184425 (2005).

\bibitem{Granados} X. Granados, J. Fontcuberta, X. Obradors, and J. B. Torrance, Phys. Rev. B \textbf{46}, 15683 (1992).

\bibitem{Devendra} D. Kumar, K. P. Rajeev, J. A. Alonso, and M. J. Mart\'inez- Lope, J. Phys: Condensed Matter \textbf{21}, 185402 (2009).

\bibitem{Zhai} H.-Y. Zhai, J. X. Ma, D. T. Gillaspie,  X. G. Zhang, T. Z. Ward, E. W. Plummer, and J. Shen, Phys. Rev. Lett. \textbf{97}, 167201 (2006).

\bibitem{Kawatani} K. Kawatani, H. Takami, T. Kanki, and H. Tanaka, Appl. Phys. Lett. \textbf{100}, 173112 (2012).

\bibitem{Sharoni} A. Sharoni, J. G. Ram\'irez, and I. K. Schuller,  Phys. Rev. Lett. \textbf{101}, 026404 (2008).

\bibitem{Wei1} J. Wei, Z. Wang, W. Chen, and D. H. Cobden, Nature Nanotechnology \textbf{4}, 420 (2009).

\bibitem{Zhang} S. Zhang, J. Y. Chou, and L. J. Lauhon, Nano Lett. \textbf{9}, 4527 (2009).

\bibitem{Wu2} J. Wu, Q. Gu, B. S. Guiton, N. P. de Leon, L. Ouyang, and H. Park, Nano Lett. \textbf{6}, 2313 (2006).

\bibitem{Cao} J. Cao, E. Ertekin, V. Srinivasan, W. Fan, S. Huang, H. Zheng, J. W. L. Yim, D. R. Khanal,  D. F. Ogletree, J. C. Grossman, and J. Wu,
Nature Nanotechnology \textbf{4}, 732 (2009).

\bibitem{Granados1} X. Granados, J. Fontcuberta, X. Obradors, L. Ma\"{n}osa, and J. B. Torrance, Phys. Rev. B \textbf{48}, 11666 (1993).

\bibitem{Medarde} M. L. Medarde, J. Phys: Condensed Matter \textbf{9}, 1679 (2009).

\bibitem{Devendra2} D. Kumar, K. P. Rajeev, A. K. Kushwaha, and R. C. Budhani, J. Appl. Phys. \textbf{108}, 063503 (2010).

\bibitem{endnote} The few small resistance jumps observed below the temperature of highest order resistance jump are due to the metal to insulator transition in crystallites not directily bridging the contact pads.

 \bibitem{McLachlan} D. S. McLachlan, J. Phys. C: Solid State Phys. \textbf{20}, 865 (1987).

\bibitem{Endnote} In our earlier report \emph{J. Phys: Condensed Matter} \textbf{21}, 185402 (2009), the supercooled region switching as a single enetity are named as 'switchable region (SR)' but the same has been called 'coherence volume' in earliier reports (\emph{Phys. Rev. B} \textbf{19}, 3580 (1979)). To keep the consistency, we have also sticked to the term 'coherence volume'.

\bibitem{Lifshitz} E. M. Lifshitz and L. P. Pitaevski, Landau and Lifshitz Course of Theoretical Physics (Physical Kinetics vol 10) (Oxford: Pergamon, 1981) chapter 12

\bibitem{Sharma} P. A. Sharma, S. El-Khatib, I. Mihut, J. B. Betts, A. Migliori, S. B. Kim, S. Guha, and S.-W. Cheong, Phys. Rev. B \textbf{78}, 134205 (2008).

\bibitem{Imry} Y. Imry and M. Wortis, Phys. Rev. B \textbf{19}, 3580 (1979).


\end{thebibliography}
\end{document}